\documentclass[twocolumn,showpacs,preprintnumbers,amsmath,amssymb]{revtex4}
\usepackage{graphicx,epsfig}% Include figure files
\usepackage{dcolumn}% Align table columns on decimal point
\usepackage{bm}% bold math
\usepackage{psfrag}
\usepackage{plain}
\usepackage{tabularx}
\usepackage[latin1]{inputenc}
\usepackage{amsmath}
\begin{document}
\draft

\title{Nontrivial temporal scaling in a Galilean stick-slip dynamics}
\author{E. J. R. Parteli$^1$, M. A. F. Gomes$^2$ and V. P. Brito$^3$}
\affiliation{1. Institut f\"ur Computerphysik, ICP, Universit\"at Stuttgart, Pfaffenwaldring 27, 70569 Stuttgart, Germany. \\ 2. Departamento de F\'{\i}sica, Universidade Federal de Pernambuco - 50670-901, Recife, PE, Brazil. \\ 3. Departamento de F\'{\i}sica, Universidade Federal do Piau\'{\i} - 64049-550, Teresina, PI, Brazil.}

\date{\today}

\begin{abstract}
We examine the stick-slip fluctuating response of a rough massive nonrotating cylinder moving on a rough inclined groove which is submitted to weak external perturbations and which is maintained well below the angle of repose. The experiments presented here, which are reminiscent of the Galileo's works with rolling  objects on inclines, have brought in the last years important new insights into the friction between surfaces in relative motion and are of relevance for earthquakes, differing from classical block-spring models by the mechanism of energy input in the system. Robust nontrivial temporal scaling laws appearing in the dynamics of this system are reported, and it is shown that the time-support where dissipation occurs approaches a statistical fractal set with a fixed value of dimension. The distribution of periods of inactivity in the intermittent motion of the cylinder is also studied and found to be closely related to the lacunarity of a random version of the classic triadic Cantor set on the line.

\end{abstract}


\maketitle

\section{Introduction}

It can be conjectured that the experimental verification of fluctuations in the dynamic behavior of a block on an incline was first verified in the beginning of the 17th century. It is known that as early as 1604 Galileo believed in the now classical scaling law $s \sim t^2$ connecting the distance $s$ and the time $t$ for an object falling close to Earth under the action of gravity \cite{ref:dugas_book}. Galileo's application of this law for {\em{rolling}} objects on inclined planes was discussed in his last book {\it{The Two New Sciences}}, published in 1638. However, no explicit or implicit mention of the application of this simple scaling law for nonrotating objects is made in Galileo's works. Certainly Galileo tried to study the pure translational motion of objects on an incline, but he was possibly unable to encapsulate the experimental results in a simple mathematical law as the previous one: the experimental outcomes in this case are very complex, apparently nonreproducible, due to the fluctuations in the friction force acting at the incline$-$sliding object interface. Few nonspecialized textbooks have called attention to the inherent fluctuating character of friction forces between solid surfaces (an exception is Ref. \cite{ref:feynman_book}). Only recently friction fluctuations involving solid surfaces have been quantitatively studied in a number of macroscopic situations \cite{ref:demirel_prl_1996,ref:briscoe_jpd_1985,ref:vallette_pre_1993}. 

In the last few years, it has been shown that intermittent sliding or stick-slip dynamics of a rough solid nonrotating cylinder on a rough inclined groove submitted to small controlled perturbations is a fluctuation phenomenon characterized by nontrivial spatiotemporal scaling laws \cite{ref:brito_pla_1995,ref:gomes_jpd_1998} and complex critical exponents \cite{ref:parteli_pa_2001} if the inclination is well below the angle of repose. In particular, the time series of intermittent slidings associated with the stick-slip motion of the cylinder on the incline were found to present many similarities to time series of earthquakes: The sliding distribution is described by the Gutenberg and Richter law $n(s)\sim s^{-0.5}$, where $n(s)$ is the number of events of size $s$, and the Omori law for the number of smaller events occurring at a time $t$ after a large event, $n(t) \sim t^{-p}$, where the exponent $p$ is an anomalous one, lying between $0.25$ and $0.45$, and may be a complex number. Stick-slip dynamics is a topic of broad interest, and it traditionally refers to the situation in which a solid on a horizontal surface is pulled at a constant driving velocity \cite{ref:persson_book}. In the experiments discussed in this paper, the stick-slip dynamics appears as a consequence of a completely different mechanism: to start the slip we resort to small mechanical perturbations on an inclined surface, whose angle with the horizontal is below the angle of repose. It is important to notice that this mechanism constitutes indeed the basic difference from many models of earthquakes, which use blocks and springs to simulate the motion of tectonic plates \cite{ref:carlson_prl_1989,ref:chen_pra_1991,ref:olami_prl_1992}. A solid body on a perturbed incline is an example of a nonequilibrium system receiving an incoming energy flux. If energy is continuously injected into nonequilibrium systems, a complex sequential response characterized by time series of events of all sizes is often observed. Besides sliding blocks on inclines, other examples of systems and phenomena associated with a similar type of temporal fluctuating response are piles of sand and other granular materials \cite{ref:held_prl_1990,ref:miller_prl_1996}; acoustic emission from volcanic rocks and microfracturing processes in general \cite{ref:diodati_prl_1991,ref:petri_prl_1994}; interface depinning in magnetic systems \cite{ref:urbach_prl_1995}; stick-slip motion in lubricated systems \cite{ref:demirel_prl_1996}; and turbulence \cite{ref:frisch_book}, among others.  
                   
This work reports on basic geometric aspects of the fluctuating behavior observed in experimental time series of sliding events of a rough nonrotating metallic cylinder moving intermittently on a rough groove weakly perturbed by external impacts and maintained well below the angle of repose. The Gutenberg \& Richter law discussed for this system in Ref.~\cite{ref:brito_pla_1995} is a relation between two {\em{macroscopic}} variables, independent of the sequence of the sliding events; furthermore, the Omori law studied in Ref.~\cite{ref:parteli_pa_2001} and the Hurst exponent analysis presented in Ref.~\cite{ref:gomes_jpd_1998} are defined for a {\em{mesoscopic}} scale associated with time windows of many slips within a sequence of sliding events. In this work, we investigate statistical variables which are strongly related to the {\em{microscopic}} details of the time series of slidings. The experiments giving origin to these time series are very time consuming but the results are quite robust in the sense that they are independent of the particular value of several experimental parameters. In particular, we learn from our experiment that (i) robust and nontrivial temporal scaling laws appear in this stick-slip dynamics, and (ii) the time-support where dissipation occurs is also statistically robust, defining a set whose dimension lies in the interval $0.61 \pm 0.01$. Moreover, (iii) the lacunarity function \cite{ref:allain_pre_1991} describing the statistical distribution of time domains free of dissipation is shown to be related to the corresponding lacunarity of a random version of the classic triadic Cantor set on the line. The {\it{Galilean}} experiments giving rise to the intermittent time series studied in this work are described in the next section. In Sec.~3 our results are presented and discussed. We then justify the use of mathematical constructions as the Cantor sets on the line to give a statistical description of the intermittent distribution of slidings on the time axis. Conclusions are made in Sec.~4.

\section{Experimental details}

\begin{figure}
  \begin{center}
    \includegraphics[width=0.95 \columnwidth]{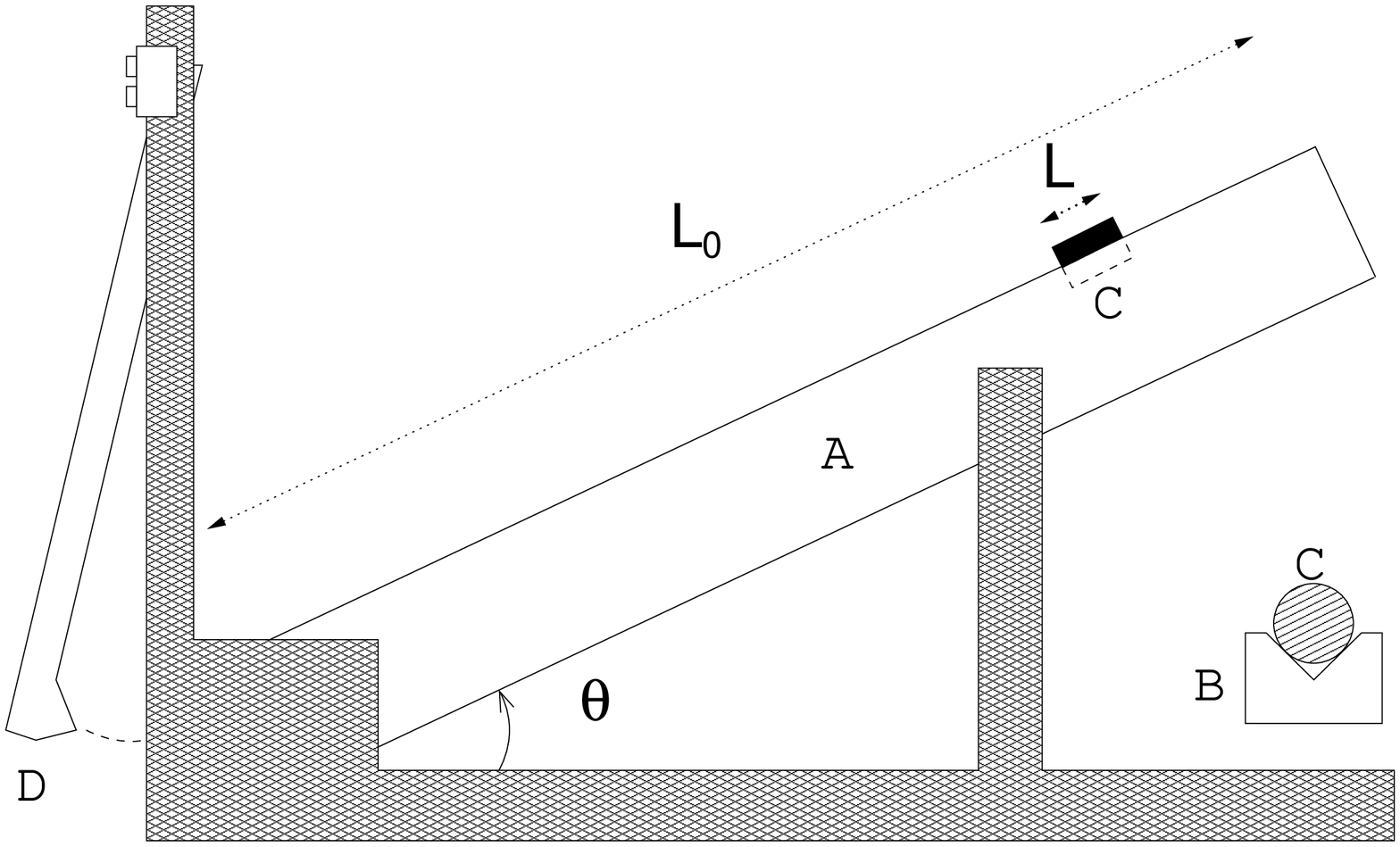} 
\caption{Schematic diagram of the experimental apparatus: the incline A has a length $L_0=1500$ mm, and is rigidly maintained at an inclination $\theta < {{\theta}_{\mathrm{c}}}$ (=angle of repose) with the horizontal. The cross section of A is shown in B; C is a cylinder with length $L$ varying from $L=5$ to $L=1000$ mm. Sliding events occur after a variable number of impacts of the hammer D. The time series discussed in the text and illustrated in Fig.~\ref{fig:time_series} are chains of 100 sliding events.}
    \label{fig:experiment}
\end{center}
\end{figure}

The basic apparatus used to obtain the experimental data discussed in this work consists of a rigid V-shaped anodized aluminum chute made of a corner plate of 5 mm thickness and with a $90^{\circ}$ angular aperture symmetrically disposed with respect to the vertical plane (Fig.~\ref{fig:experiment}). 
The chute is rigidly maintained with an inclination $\theta$ with respect to the horizontal, and it is supplied with an articulated hammer of mass $m$ which hits the base of the chute with a controlled (fixed) velocity. The results reported here are statistically invariant for $m$ in the interval studied from 75 g to 175 g. The experimental apparatus relaxes elastically within the time of 5 s between two consecutive impacts of the hammer. The system is mounted on a 200 kg table isolated from mechanical vibrations. On the chute (with an effective length $L_0=1500$ mm) is placed a metallic cylinder of length $L$ ($5\!-\!1000$ mm), and the system chute $+$ cylinder operates significantly below the critical angle of repose ${\theta}_{\mathrm{c}} = {\tan}^{-1}{\mu}_{\mathrm{s}}$, where ${\mu}_{\mathrm{s}}$ is the coefficient of static friction of the cylinder on the chute. The weak impacts of the hammer do not lead to jumps of the cylinder out of contact. In all experiments, the inclination was in the interval $12^{\circ}\!-\!18^{\circ}$, with $16^{\circ} < {\theta}_{\mathrm{c}} < 32^{\circ}$, and the reduced angle (${\theta}_{\mathrm{c}} - {\theta})/{\theta}_{\mathrm{c}}$ was typically in the interval $0.28\!-\!0.35$. 

\begin{figure}
  \begin{center}
    \includegraphics[width=1.00 \columnwidth]{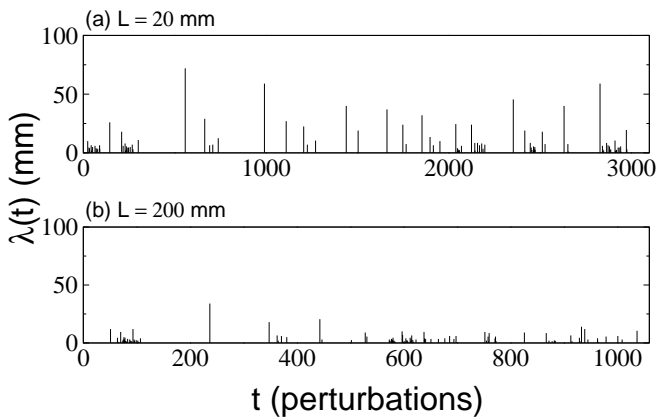} 
\caption{Typical nonhomogeneous intermittent time series of sliding events for massive cylinders of aluminum of length (a) $L = 20$ mm and (b) $L = 200$ mm, with $({\theta}_{\mathrm{c}} - {\theta})/{\theta}_{\mathrm{c}} \approx 0.30$.}
    \label{fig:time_series}
\end{center}
\end{figure}

The initial condition in all experiments refers to the cylinder at the top of the groove. Afterwards, the number of the impact of the hammer after which each sliding occurred, $t$, is recorded, as well as the corresponding sliding length ${\lambda}(t)$. In the regime of inclination used in the experiments, the series of induced sliding events are intermittent, i.e., a fluctuating number of many controlled perturbations of the hammer (time units) is necessary to induce a single sliding event of the cylinder. The total duration $T$ of each time series corresponds to the number of hammer impacts after 100 sliding events $\lambda \neq 0$, and varies from 364 to $6,979$ time units or perturbations \cite{ref:brito_2004}. The number of sliding events was fixed in such a manner that in all experiments the time series were recorded before the cylinder hit the end of the incline. Thus, each sliding sequence refers to a single sweep of the cylinder along the groove. The total number of perturbations in our experiments exceeded $56,000$, corresponding to $6,000$ sliding events. For illustration, Fig.~\ref{fig:time_series} shows typical intermittent sliding series for two values of the length $L$ of a massive aluminum cylinder of diameter $9.6$ mm.

\section{Results and discussion}

The dynamics of translation of a solid body on an inclined groove is dependent on the angle of inclination $\theta$: (i) If $\theta > {\theta}_{\mathrm{c}}$, there is a trivial single continuous sliding event (with or without perturbation on the groove). (ii) If $\theta$ is slightly smaller than ${\theta}_{\mathrm{c}}$, the continuous phase (i) is replaced by a discrete homogeneous phase with a one-to-one correspondence between perturbation and sliding response \cite{ref:brito_pla_1995}. In this case, there are spatial fluctuations in the magnitude of the sliding, but no temporal gaps between sliding events. This phase can be named Euclidean because the support of dissipation has the same dimension $D=1$ of the time axis. (iii) If $\theta$ is significantly smaller than ${\theta}_{\mathrm{c}}$ (as quantified in the previous section), there are both spatial fluctuations in the magnitude of the sliding events {\em{and}} fluctuations in the time between consecutive sliding events. (iv) For very small angles, the resting state of the cylinder is obtained. It is the nonhomogeneous intermittent regime (iii) that is studied in the present work: Fig.~\ref{fig:time_series} shows the nonhomogeneous intermittence observed in two time series obtained from our experiments. This is interesting and nontrivial because in principle the dissipation could present gaps homogeneously distributed along the time. However, this intermittent homogeneous-gap phase was never detected, i.e., nature seems to prefer the heterogeneous clustering of sliding events. Perhaps, this homogeneous gap phase could be detected in experiments using a single crystal placed on another single crystal. 

\begin{figure}
  \begin{center}
    \includegraphics[width=1.00 \columnwidth]{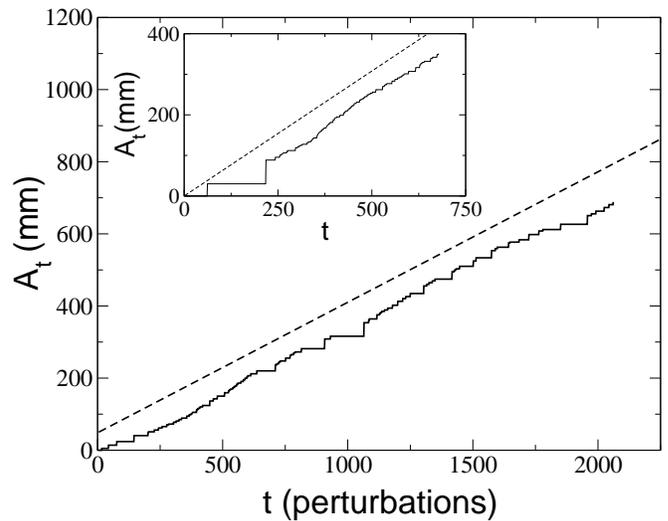} 
\caption{Cumulative sum $A_t=\sum_{i=1}^N{\lambda(t)}$ of the sliding events as a function of time for two different experiments: In the main plot, the corresponding time series was obtained with a cylinder of length $L=10$ mm, while in the inset the length of the cylinder used in the experiment was $L=100$ mm. The dashed lines in the two plots are guide to the eye meant to show the linear behavior of $A_t$.}
\label{fig:activity}
\end{center}
\end{figure}

In an experiment like the block$-$incline studied here, where many variables and parameters are involved, it is of crucial relevance to know which elements related to the apparatus are of importance for the behavior of the system, and how they could influence any finding. For example, we know that cylinders of smaller length take more time (more impacts of the hammer are needed) to slip, and this reflects the fractal nature of the roughness of the surfaces in relative motion \cite{ref:brito_2004}. Furthermore, from a ``first sight'' into our experiment, one could be easily led to the conclusion that the sliding susceptibility of the blocks should increase in time since they get closer to the impact source (the hammer) after each slip (see Fig.~\ref{fig:experiment}). However, figure \ref{fig:activity} shows that this is not the case. In the main plot of this figure, we show the cumulative sum $A_t=\sum_{i=1}^N{\lambda(t)}$ of the magnitude $\lambda$ of the sliding events as a function of time $t$ for a cylinder of length $L=10$ mm, while in the inset the same quantity is shown for a cylinder of length $L=100$ mm. As we can see from these plots, $A_t$ increases quite linearly in $t$, i.e., the sliding probability remains constant in time, and this result is independent of the length of the cylinder. In other words, the energy released by the hammer impact at the bottom of the apparatus is transmitted in average homogeneously along the chute. Figure \ref{fig:activity} has to be distinguished from Figs. $1-4$ of Ref.~\cite{ref:parteli_pa_2001}, where we plot the {\em{number}} of sliding events as a function of time, despite their magnitude. On the other side, the behavior of the system may change dramatically depending for instance on the values of the inclination of the chute and on the intensity of the hammer impact. Decreasing [increasing] the angle $\theta$ of the incline and/or the intensity of the perturbations to values far from the ranges mentioned in the last section would bring the system from the regime (iii) defined in the last paragraph to the regime (iv) [to regime (ii) or even (i)]. However, in this paper, we are just interested in the regime (iii) of nonhomogeneous intermittent motion of the cylinder on the chute. In the next paragraphs we will quantify this intermittent distribution by calculating the dimension of the dissipation support and its lacunarity.

\begin{figure}
  \begin{center}
    \includegraphics[width=1.00 \columnwidth]{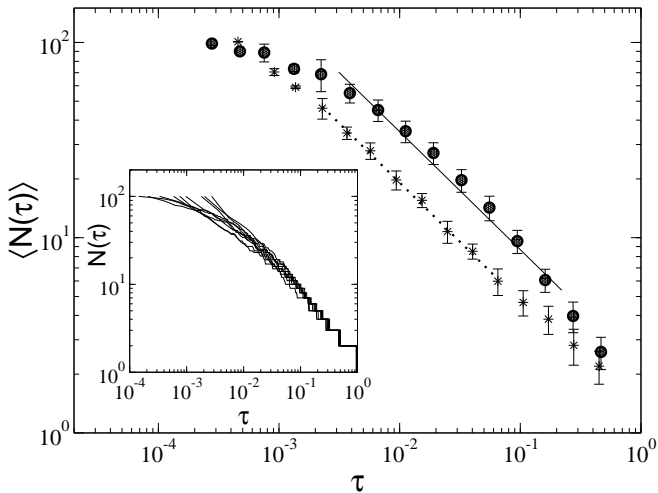} 
\caption{Log-log plot of the average number $\left<{N(\tau)}\right>$ of one-dimensional boxes of size $\tau$ necessary to cover the regions along the time axis where the sliding activity is concentrated. The average $\left<{\cdot}\right>$ is over the entire experimental set of time series of sliding events (circles). The asterisks denote the same quantity for an ensemble of equal size of synthetic series: random Cantor sets whose number of events is equal to the number (100) of sliding events (see text). The dotted line has as slope the golden mean $(\sqrt{5} - 1)/2 \simeq 0.618\ldots$. The inset shows $N(\tau)$ for typical experimental time series.}
    \label{fig:box_counting}
\end{center}
\end{figure}

After assigning ``1'' [``0''] to each time position in which the cylinder slides [is at rest], we define the temporal dissipation support $\Sigma$ as the subset of the time axis consisting of 1's. To quantify the time distribution of the sliding activity, we count the number $N(\tau)$ of one-dimensional boxes of size $\tau$ necessary to cover $\Sigma$. Here, the variable $\tau$ is defined as ${{\Delta}{t}}/T$, where ${\Delta}{t}$ is the corresponding length of the box in time units, and $T$ is the total number of perturbations. In the inset of Fig.~\ref{fig:box_counting}, we show $N(\tau)$ versus $\tau$ for some typical series. The circles shown in the main plot represent the complete ensemble average $\left<{N(\tau)}\right>$, which was obtained by dividing the $\tau$ axes into 15 equal intervals and calculating the mean values of $N(\tau)$.

For comparison, we show with asterisks in Fig.~\ref{fig:box_counting} the corresponding function $\left<{N_{\mathrm{MRCS}}(\tau)}\right>$ averaged over an ensemble of different random Cantor sets defined as follows. First, the algorithm for the triadic random Cantor set described in Ref.~\cite{ref:turcotte_book} is implemented for a line with length $T=3^7=2187$ (this value is close to the mean size of the dissipation support of the series studied, $\left<T \right> \cong 2175$): each line segment is divided into three equal parts and one of them is thereafter randomly removed. After the 7th iteration, this random construction contains just $2^7 = 128$ events. We then take out 28 events randomly, to get a modified random Cantor set (MRCS) with the same number (100) of events as in the experimental series. $\left<{N_{\mathrm{MRCS}}(\tau)}\right>$ scales as ${\tau}^{-D_{\mathrm{MRCS}}}$, with $D_{\mathrm{MRCS}} = 0.61 \pm 0.01$, as shown by the dotted line in Fig.~\ref{fig:box_counting}. The continuous line in the plot has the same slope as the dotted one. As we can see, this line gives a good approximation to the corresponding slope of the experimental trend line along almost two decades in $\tau$. Although the experimental data clearly do not present a perfect scaling - due to the low dimensionality of the phenomenon and the consequent difficulty to obtain series with many events $-$ $\left<{N(\tau)}\right>$ indicates {\em{within the error bars}} that the dissipation support {\em{approaches}} a set with dimension $D_{\mathrm{MRCS}}=0.61 \pm 0.01$, which is the fractal dimension of the random Cantor set defined above. It is important to notice that the scaling relation shown in Fig.~\ref{fig:box_counting} does not mean that the dissipation support {\it{is}} a fractal set. Moreover, all the random Cantor sets studied here are {\it{statistical}} fractal sets and have in average the same finite size $\left<{T}\right>$ of the experimental series, as well as the same number of events as in the corresponding sliding sequences. A quite general random Cantor set construction was introduced by Mauldin and Williams \cite{ref:mauldin_tams_1986}. It consists of the recursive division of the line segment into three parts of different sizes, which are determined by two random parameters $0 < r_1, r_2 < 1$, followed by the removal of the central segment. The dimension of the random Cantor sets obtained with this algorithm is known as the golden mean, $D_{\mathrm{MW}} \equiv {(\sqrt{5} - 1)}/2 = 0.618 \ldots$. Mauldin and Williams showed that although the infinite combinations of $r_1$ and $r_2$ lead to completely different geometries, the resulting random Cantor sets appear robust with respect to their dimension. The value of $D_{\mathrm{MW}}$ is universal and close to the value $D_{\mathrm{MRCS}}=0.61 \pm 0.01$ obtained from Fig.~\ref{fig:box_counting}. We speculate that this general class of random recursive dynamics introduced by these authors could be associated with the physical phenomenon discussed in the present work.

\begin{figure}
  \begin{center}
    \includegraphics[width=1.00 \columnwidth]{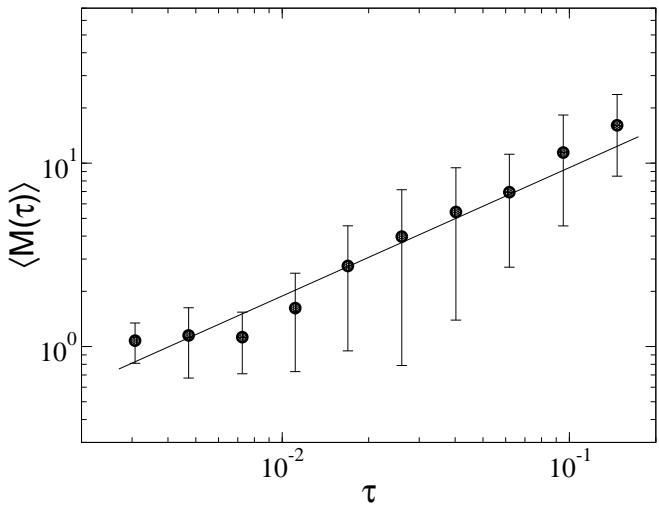} 
\caption{Log-log plot of the average number $\left<{M(\tau)}\right>$ of sliding events within an interval of time $\tau$. The average $\left<{.}\right>$ is over the entire experimental set of time series of sliding events (circles). The line has slope $\delta = 0.65 \pm 0.05$, which is interpreted here as the (mass) dimension of the statistic fractal set defined by the dissipation support of the experimental time series. This value is in agreement, within the fluctuation bars, with the dimension found in Fig.~\ref{fig:box_counting} calculated with the box counting for the same values of the time interval $\tau$.}
    \label{fig:mass_dimension}
\end{center}
\end{figure}

An additional method to define a fractal dimension is based on the mass-size relation \cite{ref:mandelbrot_book}. In our case this means to count the number of sliding events, $M(\tau)$, within an interval of time $\tau$. For a scaling distribution of intermittent events on the time axis we expect that the ensemble average $\left<{M(\tau)}\right> \sim {\tau}^{\delta}$, where the mass-exponent ${\delta}<1$. In Fig.~\ref{fig:mass_dimension} we show that, in fact, $\left<{M(\tau)}\right>$ behaves as a power law of $\tau$ with the exponent $\delta = 0.65 \pm 0.05$, which is essentially the same as $D_{\mathrm{MRCS}}$ within the fluctuation bars. Again, this figure has to be distinguished from Fig.~3 in Ref.~\cite{ref:parteli_pa_2001}, where we show the number of events {\em{between two large slips}}, which provides a {\em{mesoscopic}} description of the time series, as opposed to the variable $\left<{M(\tau)}\right>$, which quantifies the {\em{microscopic}} distribution of sliding events along the dissipation support.

Another way to get randomized Cantor sets in our problem would be taking out randomly 28 of the 128 events of a seventh order triadic Cantor set (TCS), whose fractal dimension is $D_{{\mathrm{TCS}}} = {\ln{2}}/{\ln{3}} = 0.6309\ldots$ (obtained by recursive removal of the central third of each segment \cite{ref:mandelbrot_book,ref:turcotte_book}). Let us define a modified triadic Cantor set (MTCS) of seventh order as a triadic Cantor set after the random removal of 28 events. It can be easily verified that this operation leads to a set with dimension $D_{\mathrm{MTCS}} = 0.61 \pm 0.01$. This means that both the MRCS and the MTCS simulate, from the point of view of fractal dimension, the dissipation support of the time series of sliding events. However, it is clear that these two kinds of Cantor sets are very different in respect to their origin. Clearly, the MRCS is more random than the MTCS. To remove the degeneracy between these two sets due to their common value of fractal dimension, we examine another statistical property: the lacunarity function.

\begin{figure}
  \begin{center}
    \includegraphics[width=1.00 \columnwidth]{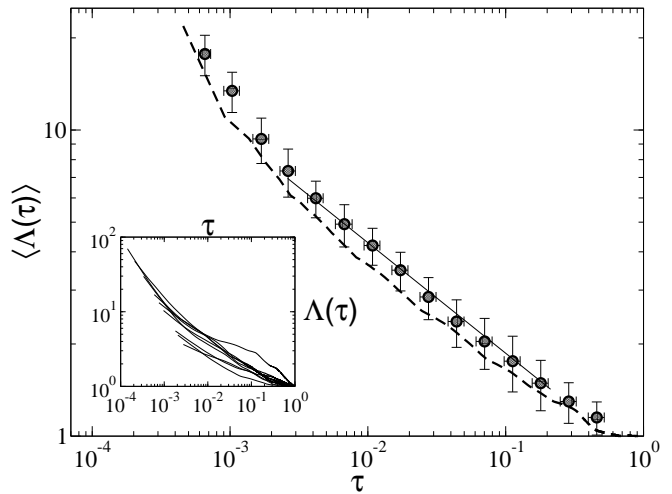} 
\caption{Log-log plot of the lacunarity function $\left<{{\Lambda}({\tau})}\right>$ for the experimental time series (circles). The dashed line gives the same function for an ensemble of MTCS of order 7 ($2^7$ events, $T = 3^7$) with 28 events randomly eliminated (see text). The inset shows ${\Lambda}({\tau})$ for some typical time series. $\left<{{\Lambda}({\tau})}\right>$ decays effectively as the power law ${\tau}^{-0.36}$ for $3 \times 10^{-3} \leq {\tau} \leq 2 \times 10^{-1}$ (continuous line).}
    \label{fig:lacunarity}
\end{center}
\end{figure}

Lacunarity, a concept introduced by Mandelbrot, tries to quantify the texture of a fractal \cite{ref:mandelbrot_book}. Here we use the mathematical algorithm introduced by Allain and Cloitre \cite{ref:allain_pre_1991} to calculate the lacunarity of the dissipation support of the sliding events. A short description of their method can be given as follows. First, a time window of size $\tau$ is defined, which is translated by $R(\tau) = T{(1-\tau)} + 1$ steps (time units), from the beginning up to the end of the time series, which has size $T$. The number $n(s,{\tau})$ is defined as the number of times in which the window of size $\tau$ contains $s$ sliding events during translation. Thus the probability that the window encounters $s$ occupied time units during translation is $p(s,{\tau}) = {n(s,{\tau})}/{R({\tau})}$. It follows that the lacunarity ${\Lambda}(\tau)$ can be defined as
\begin{equation}
{\Lambda}(\tau) = \frac{{\sum_{s=1}^{\tau}}{s^2}\,p(s,\tau)}{{\left[{{\sum_{s=1}^{\tau}}{s\,p(s,\tau)}}\right]}^2} \: .
\end{equation}
This function has very interesting properties, being equal to unity when $\tau=1$. The inset of Fig.~\ref{fig:lacunarity} shows the lacunarity $\Lambda({\tau})$ calculated for some time series studied. In the main plot we show the experimental ensemble average value $\left<{{\Lambda}({\tau})}\right>$, represented by circles, together with the corresponding fluctuation bars, as a function of ${{\tau}}$. As we can see from this figure, the dashed line follows very closely the trend of the curve defined by the circles. This line is no best fit obtained from the experimental data: it is the lacunarity calculated for the MTCS of seventh order of the previous paragraph. The agreement of the curves is noticeable. It can be seen from Fig.~\ref{fig:lacunarity} that $\left<{{\Lambda}({\tau})}\right>$ decays effectively as the power law ${{\tau}}^{-0.36}$, along approximately two decades of variability in time, for $3 \times 10^{-3} \leq {\tau} \leq 2 \times 10^{-1}$, as indicated by the continuous line.

The Cantor sets discussed here are mathematical structures with topological dimension $d_{\mathrm{t}}=0$, fractal dimension $D$ in the interval $(0,1)$, and embedding dimension $d=1$ \cite{ref:mandelbrot_book}. As shown in Fig.~\ref{fig:time_series}, intermittence does not distribute homogeneously on the time axis, i.e., its temporal support may, in principle, be characterized by a dimension $D<1$. Thus Cantor sets are natural candidates to model nonhomogeneous intermittence and other fluctuation phenomena defined along the 1D time axis. Brownian motion is the prototype of the fluctuation phenomenon: if we plot the position $X$ of a particle executing Brownian motion in one-dimensional (1D) as a function of the time $t$, we obtain a fractal record $X(t)$ having dimension $D_B =3/2$ on the position-time plane. The zeroset for the Brown function $X(t)$, which is defined as those instants $t$ for which $X(t)$ cuts the time axis, is also a Cantor set, but one with dimension $D=D_B-1=(3/2)-1=0.5$ \cite{ref:mandelbrot_book}. Analogously, the temporal dissipation support studied in the present work can be seen as a zeroset associated with the sliding length distribution ${\lambda}(t)$, albeit a zeroset with larger (and equally robust, as in the Brownian case) dimension in the interval $0.61\!-\!0.65$.

Finally, we must ask why indeed should this type of experiment exhibit nonhomogeneous intermittence with a random response of sliding events? A possible explanation is that the number and the individual characteristics of the several contacts at the cylinder$-$chute interface fluctuate continuously on the time scale associated with the hammer impacts. Eventually, e.g., a single interface contact providing a force equilibrating a large (small) fraction of the component of the cylinder's weight along the chute fails and, as a consequence, a corresponding large (small) resulting force appears, with the cylinder tending to stop after a long (short) sliding length. The motion of the cylinder is followed by the rest state when the fluctuating interfacial roughness is able to provide a pattern of contacts producing the necessary opposing force to equilibrate the weight's component down the chute. Evidently, as the angle $\theta$ (or the component of the cylinder's weight down the chute) decreases, it becomes on average increasingly difficult to modify the rest state of the cylinder; i.e., the sliding susceptibility decreases \cite{ref:brito_2004}. The observed intermittence with fluctuating gaps denoting the interval between sliding events (dissipation) would be a manifestation of the fluctuation in the ability of the cylinder$-$chute contacts in maintaining the equilibrium of forces for the cylinder. This situation is reminiscent of the fluctuation-dissipation theorem: If the contacts and the corresponding forces at the interface cylinder$-$chute do not fluctuate, the cylinder remains at rest, i.e., there is no dissipation or sliding event; on the contrary, for increasing fluctuations at the interfacial forces, the dissipation also increases, i.e., a cascade of intermittent sliding events is obtained as the associated response.

\section{Conclusion}

We found robust nontrivial temporal scaling laws related with the complex time-distribution of sliding events observed in an extensive experimental study of Galilean stick-slip dynamics on rough inclined surfaces. It is found that the time-support where dissipation occurs in these processes approaches a statistical fractal set characterized by a fixed dimension, $D=0.61 \pm 0.01$. The lacunarity function for the time series of events is closely related to the corresponding lacunarity of a modified random version of the classic triadic Cantor set \cite{ref:mandelbrot_book}. It has to be emphasized that the geometric and statistic properties exhibited by the time series of sliding events are a consequence of the particular nature of the microscopic friction at the block$-$incline interface. No spatial and temporal correlations as shown here and as those reported in \cite{ref:brito_pla_1995,ref:gomes_jpd_1998,ref:parteli_pa_2001,ref:brito_2004} are found in these series if the block is moved back to the top of the incline after each sliding event, which would mean removing the strong correlation between the configuration of the contacts at the interface and the past history of the block on the incline. Due to the very time-consuming nature of the experiments described in this paper, many aspects remain to be investigated in the future. It is important to notice that the results presented in this paper are restricted to time series obtained from the intermittent stick-slip motion of the cylinder, which means the regime (iii) defined in section 3, and thus constrains the values of experimental parameters as, for instance, the inclination $\theta$ of the chute and the mass of the hammer to the ones mentioned in section 2, as said before. However, new experiments exploring systematically the effect of the intensity of the perturbation on the groove, as well as the variation of the materials involved and the effect of the geometry of the sliding objects are necessary for a better understanding of this fluctuation phenomenon. Furthermore, another important aspect to be considered is the possibility of a universal value $D = 0.61 \pm 0.01$ for the dimension of the dissipation support in any experiment involving sliding of rough solids of the type discussed in this paper. In a recent paper \cite{ref:feng}, Feng and Seto have analyzed several time series of acoustic emissions obtained from microfracturing in rocks, which is a process intimately connected with the dynamics of earthquakes, and have found that triggering events associated with a failure define a dissipation support with values of dimension close to the exponents found for our time series of slidings. The application of the analysis reported in the present work to the temporal response of other nonequilibrium systems as those mentioned in the end of the second paragraph could clarify important aspects of universality in nonequilibrium dissipative systems.

\acknowledgments
We acknowledge M. Lyra, H. J. Herrmann, P. G. Lind and F. Raischel for stimulating remarks and helpful comments. We are also grateful to V. Schatz for a critical reading of this manuscript. M. A. F. G. acknowledges financial support from CNPq and PRONEX (brazilian agencies). E. J. R. Parteli acknowledges support from CAPES - Bras\'{\i}lia/Brazil.

\end{document}